\documentstyle[11pt,aaspp4]{article} \parskip 0pt

\begin{document}

\let\mic=\micron
\let\deg=\arcdeg
\def\Ref{\reference{}}
\def\eq#1{\begin{equation} #1 \end{equation}}
\def\E#1{\hbox{$10^{#1}$}}
\def\about  {\hbox{$\sim$}}
\def\ga     {\hbox{$\gtrsim$}}
\def\la     {\hbox{$\lesssim$}}
\def\x      {\hbox{$\times$}}
\def\T      {\hbox{$T_\ast$}}
\def\R      {\hbox{$R_\ast$}}
\def\Rd     {\hbox{$R_{\rm disk}$}}
\def\tV     {\hbox{$\tau_V$}}
\def\Tdout  {\hbox{$T_{\rm disk}^{\rm out}$}}
\def\tD     {\hbox{$\tau^{350}_{\rm disk}$}}
\def\Rsub   {\hbox{$R_{\rm sub}$}}
\def\Tsub   {\hbox{$T_{\rm sub}$}}
\def\Lo     {\hbox{$L_\odot$}}
\def\Mo     {\hbox{$M_\odot$}}
\def\Ro     {\hbox{$R_\odot$}}
\def\lFl    {\hbox{$\lambda F_\lambda$}}
\def\f      {\hbox{$f_\lambda$}}
\def\Fdisk  {\hbox{$F_{\rm disk}$}}
\def\fdisk  {\hbox{$f_{\rm disk,\lambda}$}}
\def\Fenv   {\hbox{$F_{\rm env}$}}
\def\fenv   {\hbox{$f_{\rm env,\lambda}$}}

\def\Msg{Submitted April 13, 1999; accepted May 26, 1999}
\rightline{ApJ Letters, \Msg}

\title{DUST EMISSION FROM HERBIG Ae/Be STARS ---
    \\ EVIDENCE FOR DISKS AND ENVELOPES}

\author{Anatoly Miroshnichenko\altaffilmark{1},
        \v{Z}eljko Ivezi\'c\altaffilmark{2},
        Dejan Vinkovi\'c\altaffilmark{3}
        and Moshe Elitzur\altaffilmark{3,4}}

\altaffiltext{1} {Pulkovo Observatory, St. Petersburg 196140, Russia;
                  anat@pulkovo.spb.su}
\altaffiltext{2} {Princeton University, Department of Astrophysical Sciences,
                  Princeton, NJ 08544; ivezic@astro.Princeton.edu}
\altaffiltext{3} {Department of Physics and Astronomy, University of Kentucky,
                  Lexington, KY 40506; dejan@pa.uky.edu, moshe@pa.uky.edu}
\altaffiltext{4} {Service d'Astrophysique, CEA Saclay, 91190 Gif-sur-Yvette,
                  France}

\begin{abstract}

IR and mm-wave emission from Herbig Ae/Be stars has produced conflicting
conclusions regarding the dust geometry in these objects. We show that the
compact dimensions of the mm-wave emitting regions are a decisive indication
for disks. But a disk cannot explain the spectral energy distribution (SED)
unless it is embedded in an extended envelope that (1) dominates the IR
emission and (2) provides additional disk heating on top of the direct stellar
radiation.  Detailed radiative transfer calculations based on the simplest
model for envelope-embedded disks successfully fit the data from UV to mm
wavelengths and show that the disks have central holes. This model also
resolves naturally some puzzling results of IR imaging.

\end{abstract}


\section{INTRODUCTION}

Attempts at determining the dust geometry around Herbig Ae/Be stars (HAEBES)
from IR and mm-wave data have yielded conflicting conclusions.  From the shape
of the SED, Hillenbrand et al.\ (1992) suggested that the IR emission from 30
HAEBES is dominated by optically thick accretion disks. However,
Miroshnichenko, Ivezi\'c \& Elitzur (1997; MIE) successfully modeled the
detailed data of eight of these objects with optically thin spherical envelopes
in free fall. Worse yet, occasionally the data at different wavelengths from
the same source seem in conflict.  Mannings \& Sargent (1997; MS) measured the
mm-wave emission from seven HAEBES, including two of the MIE sources. They find
that the visual optical depths (\tV) required to explain the mm emission with
the MIE models are at least 220, at considerable odds with the small $V$-band
extinction of each source. In particular, MIE successfully fitted all $\lambda
\le$ 100 \mic\ data with \tV\ = 0.4 for MWC 480 and \tV\ = 0.3 for MWC 863, yet
MS find that within the context of the MIE model, the mm-emission from these
sources requires $\tV > \E3$ and \tV\ = 601, respectively. However, Mannings
and Sargent's proposed solution that the emission originates from
optically-thick accretion disks yields similar inconsistencies. In this case
$F_\nu \propto \nu^{1/3}$, and extrapolating the flux from the MS mm
measurements produces IR emission that is more than an order of magnitude too
weak. For example, for MWC 863 the measured 2.6 mm flux of 13.7 mJy
extrapolates to only 0.1 Jy at 2.2 \mic, where the observed flux is 4.6 Jy.
Also, the proposed accretion disks would be optically thick at 10 \mic, where
the silicate feature indicates prominent optically thin emission in most of the
MS sources.

Similar inconsistencies arise from the HAEBES imaging by Di Francesco et al.\
(1994; DF). Irrespective of geometry, the longer the radiation wavelength, the
cooler the emitting region is, and since temperature drops with distance from
the center, the image size should increase with wavelength. While this was the
case for most sources, the image size of MWC 137 was $66\arcsec \pm 2\arcsec$
at 50 \mic\ and only $58\arcsec \pm 2\arcsec$ at 100 \mic. No single dust
configuration can produce a decrease of observed size with wavelength.

We propose a simple solution to the internal inconsistencies that seem to
afflict the observations at different wavelengths of some HAEBES: the dust
distribution in these sources has both an extended spherical component,
dominating the IR emission, and an embedded compact disk which dominates the mm
and sub-mm emission. Here we show that this simple model resolves all the
conflicts quite naturally.

\section{MODELING AND RESULTS}

The system consists of a star of radius \R\ and effective temperature \T,
surrounded by a geometrically thin and optically thick passive disk extending
from \R\ to some outer radius \Rd.  In addition, a spherical dusty envelope
starts at the radius \Rsub\ corresponding to dust sublimation.  We have
analyzed this system with the aid of the scaling theory of Ivezi\'c \& Elitzur
(1997; IE) and the classical accretion disk theory as adapted to T Tau stars by
Bertout, Basri \& Bouvier (1988). Details of our analysis will be reported
elsewhere. Here we present detailed model calculations, performed with the code
DUSTY (Ivezi\'c, Nenkova \& Elitzur 1997), that successfully fit the SEDs of
all the stars in the MS sample.

Our modeling procedure is similar to MIE except that each model flux is the sum
of disk and envelope contributions, where the latter includes also the
attenuated stellar emission. For any flux distribution $F_\lambda$ introduce
the dimensionless, normalized SED $\f = \lFl/\!\int\!F_\lambda d\lambda$, which
depends only on dimensionless quantities --- luminosities, densities and linear
dimensions are irrelevant (IE).  The only relevant property of the stellar
radiation is its spectral shape, taken from the appropriate Kurucz (1994) model
atmosphere. For the dust, the only relevant properties are the spectral shapes
of the absorption and scattering coefficients, which we take from standard
interstellar mix, and the sublimation temperature, which we take as \Tsub\ =
1500~K. DUSTY performs a self-consistent calculation of the temperature
profiles\footnote{There is no need to consider stochastic heating of very small
grains.  The stellar radiation field is sufficiently intense that all grains
are in thermal equilibrium with it (Jones 1999).} of the disk and the envelope,
taking into account both the scattered and attenuated stellar radiation and
diffuse envelope emission; the effect of disk emission on the envelope
temperature is negligible for the parameters considered here. The envelope SED,
\fenv, is determined by the envelope optical depth, \tV, and the dimensionless
profile of its density distribution in terms of $y = r/\Rsub$, where $r$ is
distance from the star. Here we employ the simple profile $y^{-p}$, with $p$ as
a free parameter, extending from $y = 1$ to some $y = Y$. If shadowing by the
star is neglected then $\Fdisk_{,\lambda} \propto \cos i$, where $i$ is the
disk inclination angle, and the disk SED, \fdisk, is independent of $i$. The
disk is assumed to be optically thick everywhere at the peak of the Planckian
with the local temperature. Then \fdisk\ has only two free parameters --- the
temperature and normal optical depth of the disk outer edge, which we denote
\Tdout\ and \tD, respectively, the latter specified at 350 \mic. Once \fdisk\
and \fenv\ are computed, the observed SED is fitted through $\f = \rho\fdisk +
(1 - \rho)\fenv$, where $\rho = \Fdisk/(\Fdisk + \Fenv)$ is a free parameter.
The final free parameter is $A_V$, the interstellar extinction to the system.

%

Figure 1 shows our modeling results, including all available data from 1400
\AA\ to 2.7 mm.  In addition to the IRAS LRS data when available, the plot for
each object contains 20--30 data points from various sources. The model
parameters obtained for each star are listed in Table 1. It is worth noting
that the $A_V$ we find are similar to those derived by MS. As is evident from
the figure, envelopes with a simple power law density distribution adequately
fit almost all sources. In all those cases the figure displays the model with
$Y = 1000$, but there is considerable freedom in this parameter and successful
models can be constructed with $Y$ as small as \about\ 150. In addition,
acceptable fits can be obtained when the power $p$ is reduced by as much as 0.5
in most cases. The largest freedom exists for MWC 758 which lacks LRS data, and
we present the two models that bracket the range of $p$; any value in between
is possible. The one exception is AB Aur, this sample's most luminous object,
which requires an extended envelope ($Y$ = 5000) with a broken power law
density profile. Indeed, it is the only source in this sample surrounded by a
visible nebulosity (Herbig 1960); the nebulosity size (\about\ 1.5 arcmin)
agrees with our model requirements. A flattening of the density distribution
away from the center can be expected at the late evolutionary stages of
collapsing clouds (e.g., Shematovich, Shustov \& Wiebe, 1997).

The key to the resolution of the conflicts outlined above is the great
disparity between the disk and envelope temperature profiles. While heating by
stellar radiation produces disk temperature that varies as $r^{-3/4}$, the
envelope temperature decreases only as $r^{-0.36}$ for dust opacity $\propto
\lambda^{-1.5}$. As a result, the disk is much cooler than the envelope at all
radii at which both exist and can also contain cooler material in spite of
being much smaller.  Both properties are evident from the top panel of figure
2, which shows the temperature profiles of the two components in AB Aur. Natta
(1993) pointed out that heating by the spherical dusty envelope significantly
affects the disk temperature, and our calculations confirm this important
point.  In figure 1, the first bump (around 1 \mic) in each disk emission
reflects the stellar heating, the second is produced by the envelope heating.
But this does not alter the fundamental difference between the temperature
profiles of the two components, as the AB Aur case shows. Although it is more
compact, the disk can be the stronger emitter at long wavelengths so that the
SED is dominated by the envelope at IR wavelengths and by the disk at mm
wavelengths. This is the case for all sources in fig.\ 1.

This role reversal affects also the wavelength behavior of images. At shorter
wavelengths the image is dominated by the envelope, and the observed size
increases with wavelength. When the SED switches to disk domination, the
observed size can decrease because a given temperature occurs on the disk at a
much smaller radius than in the envelope. This effect is evident in figure 2,
which shows the surface brightness profiles of the AB Aur model at various
wavelengths.  These profiles agree well with imaging observations (DF, Marsh et
al.\ 1995, MS).  The finite beam size and dynamic range of any given telescope
could result in an apparent size decrease between 10 \mic\ and 100 \mic\ in
this case. A switch from envelope to disk domination provides a simple
explanation for the otherwise puzzling decrease in the observed size of MWC 137
between 50 \mic\ and 100 \mic. A similar effect was recently detected also in
the dust-shrouded main-sequence star Vega. Van der Bliek, Prusti \& Waters
(1994) find that its 60 \mic\ size is $35\arcsec \pm 5\arcsec$, yet 850 \mic\
imaging by Holland et al.\ (1998) produced a size of only $24\x21\arcsec \pm
3\arcsec$. So the dust distribution around Vega, too, could have both spherical
and disk components.

%

\section{DISCUSSION}

Our models successfully fit the entire MS sample, resolving all the earlier
discrepancies.  Both disk and envelope are crucial components.  A purely
spherical distribution could successfully fit each SED, but the cool mm-wave
emitting material would have to be placed about 100 times further from the star
than the MS observations indicate (an example is the recent spherical fit for
AB Aur by Henning et al., 1998).  The compact mm-wave emission observed from
these sources can be produced only by the ``classic" geometrically-thin
optically-thick disks, as correctly recognized by Mannings \& Sargent. But such
disks alone are incapable of explaining the observations. This is evident from
figure 1, which shows the maximum possible emission from this configuration,
obtained when a ``bare" disk is observed face-on. In all sources this emission
falls short of observations at $\lambda$ \ga\ 5 \mic, mostly by substantial
amounts. The envelope is essential not only for its direct IR flux which
dominates the observations at these wavelengths, but also for its indirect
effect on the sub-mm and mm-wave emission, which is disk dominated; the
observations at these wavelengths cannot be explained without the additional
disk heating by the envelope.

Our detailed model results depend on the simplifying assumptions, but the main
conclusions seem robust: the density distribution contains two distinct
components, one optically thick, cool and compact, the other optically thin,
warmer and more extended. The spherical idealization is not essential for the
latter since the envelopes can be flattened and even distorted into irregular
shapes before severely affecting the results. Recently Chiang \& Goldreich
(1997; CG) pointed out that the optically thin emission from the surface layer
of an optically thick disk can significantly affect the SED, and in principle
this layer could fulfil the role of the envelope advocated here. However, the
emission from the CG layer can be shown equivalent to that from a spherical
envelope with optical depth \tV\ = 0.8\R/\Rsub\ and density profile $p = 2$ for
$y < 6$ and $p = 5/7$ thereafter. Detailed model calculations with these
equivalent envelopes show that they cannot fit the MS sources. In particular,
all the MS sources require $\tV > 0.1$ for their observed fluxes but only have
\R/\Rsub\ \about\ 0.01 (Table 1); i.e., the column of optically thin dust
contained in the disk surface layer is only \about\ 10\% of what the
observations require, and we are justified in neglecting this layer in our
HAEBES model calculations. On the other hand, in T Tau stars, the subject of
the CG study, \R/\Rsub\ is an order of magnitude larger than for HAEBES
(because \T\ is lower) and the surface layer becomes a significant component.
The MS sources HD 245185 and MWC 758 are potential exceptions because good fits
to their SEDs are possible with rather flat density profiles and modest optical
depths. Such envelopes could be made equivalent to CG layers if the parameters
of the CG model are scaled to the HAEBES environment keeping the basic
assumptions intact. Settling the issue with certainty for these two cases
requires a 2D radiative transfer code, which we are currently developing. In
all other sources, our conclusion about the negligible role of the disk surface
layer seems secure.

Almost all the envelopes have $\tV < 1$, therefore their material is largely
atomic.  With standard dust abundance, the envelope column densities are
\about\ \E{20} cm$^{-2}$.  All stars in this sample have \R\ \about\ 2\Ro,
therefore \Rsub\ \about\ \E{13} cm and the densities at the envelope inner
regions are \about\ \E7 cm$^{-3}$. In contrast with MS, the DF selection
criterion was high luminosities. Since \Rsub\ scales with $L^{1/2}$, the DF
sources should have more extended envelopes, as observed. The envelope mass
strongly depends on its outer radius, and this parameter is rather poorly
constrained. With $Y$ = 1000, which was employed in the displayed fits,
envelope masses range from \about\ \E{-4} \Mo\ for the sources with $p < 2$ to
\about\ \E{-6} \Mo\ for the sources with $p = 2$. However, these mass estimates
decrease sharply for smaller values of $Y$, which are possible in all cases.
The one exception is AB Aur, where there is little freedom in the outer radius,
and the model parameters give an envelope mass of 0.03 \Mo, same as a recent
estimate by Henning et al (1998). The power $p$ = 2 could indicate outflow with
constant velocity; indeed, N~V emission from AB Aur was recently modeled with a
wind (Bouret, Catala \& Simon 1997). However, acceptable fits can be produced
also with $p$ = 3/2. If this index is interpreted as steady-state accretion to
a central mass, the envelope optical depths translate to accretion rates of
\about\ \E{-8} \Mo~yr$^{-1}$, similar to those deduced from UV spectra of
HAEBES (Grady et al. 1996) and T Tau stars (Valenti, Basri \& Johns 1993;
Gullbring et al 1998). These low rates cannot correspond to the main accretion
buildup of the star but rather a much later phase, involving small, residual
accretion from the environment. The corresponding accretion luminosities are
only \about\ 0.1 \Lo, justifying their neglect in our calculations.

Since the disk is optically thick in our model, its density distribution
remains undetermined and we cannot improve on the MS estimates of disk masses
(\about\ \E{-2} \Mo).  Useful information can be deduced from the parameter
$\rho$ because it is easy to show that $\rho = 2x\cos i /(1 - x + 2x\cos i)$,
where $x$ is the disk fractional contribution to the overall luminosity.  Our
calculations automatically determine $x$ in each case, allowing us to deduce
$i$ from the model fit for $\rho$. If the disk extends all the way to the
stellar surface then $x$ = 0.4 for the AB Aur model\footnote{A ``bare" disk
that extends to the stellar surface has $x$ = 0.25.} which translates to $i$ =
80\deg\ for this source, similar to the 76\deg\ that MS deduced from the disk
elliptical appearance. However, following the same procedure for all other
stars produces $i > 80\deg$ in each case. Since an edge-on orientation for
every single disk in this sample is highly unlikely we conclude that $x$ cannot
be as large as this procedure implies for all the systems. Indeed, central
holes would drastically reduce $x$ because of the steep dependence of disk
luminosity on the radius of the disk inner edge. Moving this edge from \R\ to
only 2\R\ removes 56\% of the stellar luminosity intercepted by the disk, 3\R\
results in a 72\% removal. Central holes of virtually any size would sharply
decrease $x$, resulting in a more plausible distribution of inclination angles.
Such holes would not impact any other model result because they remove only the
hottest disk material whose contribution to the observed flux is negligible in
all cases. The sizes of these holes cannot be determined from the modeling, but
their existence seems an unavoidable conclusion.

The discrepancies among previous HAEBES studies underscore the importance of
combining multi-wavelength data in an integrative approach.  Single wavelength
observations, however detailed, never fully reconstruct the geometry of dust
distribution. At $\lambda$ \la\ 3 \mic\ the observed radiation is scattering
dominated because dust emission would require temperatures higher than
sublimation. Scattered photons trace the density distribution, so images at
these wavelengths reveal the actual geometry --- but only close to the center,
to scattering optical depth \about\ 1.  In contrast, radiation at longer
wavelengths can map much farther regions because it is dominated by dust
emission.  However, the emission is predominantly governed by the dust
temperature distribution, which primarily reflects distance from the central
star and thus tends to be spherically symmetric even when the density
distribution is not. By example, images of the nebulosity around the late-type
star IRC+10216 are elongated at $\lambda$ \la\ 3--4 \mic\ yet spherically
symmetric at longer wavelengths (Ivezi\'c \& Elitzur 1996). Recent NICMOS
images of young stellar objects show complex morphologies (Padgett et al.\
1999), and there is no reason to believe they should be simpler for HAEBES.
Nevertheless, resolving such details does not alter the measured SED. The model
parameters deduced here can be expected to provide a reasonable description of
the envelope properties when small-scale structure is averaged out.

\acknowledgments

We would like to thank Dr.\ A. Jones for useful discussions of small grain
heating. The partial support of NASA grants NAG 5-3010 and NAG 5-7031 is
gratefully acknowledged.


\def\h#1{\colhead{#1}}
\def\aD{$\theta_{\rm disk}$}
\def\d{$^\dag$}

\begin{deluxetable}{lcrcrrcrcc}
\tablewidth{0pc}
\tablecaption{Properties of Modeled Systems}
\tablehead{
\h{Name} & Sp.T & \h{$A_V$} & \h{$\rho$} & \h{\tV} & \h{$p$} & \h{\Tdout}&
\h{\tD}  & \h{\aD}   & \h{${\Rsub\over\R}$}
\\
         & \h{(1)}  & \h{(2)}  & \h{(3)}    & \h{(4)} & \h{(5)} & \h{(6)}&
\h{(7)}  & \h{(8)}  & \h{(9)}
}
\startdata
AB Aur\d   & A0 & 0.2 & 0.21 & 0.5  & 2\d & 25 & 0.9     & 2.5 & 98  \\
MWC 480    & A3 & 0.4 & 0.14 & 0.25 & 2   & 20 & 18      & 2.0 & 88  \\
HD 245185  & A0 & 0   & 0.14 & 0.6  & 1   & 25 & 2       & 1.8 & 91  \\
CQ Tau     & A8 & 0.1 & 0.17 & 2.7  & 1   & 30 & 5       & 2.6 & 79  \\
MWC 863    & A3 & 1.2 & 0.05 & 0.45 & 2   & 40 & $\gg$20 & 1.0 & 90  \\
HD 163296  & A3 & 0.3 & 0.17 & 0.3  & 2   & 20 & $>$20   & 3.1 & 94  \\
MWC 758(A) & A8 & 0.2 & 0.49 & 0.25 & 0.5 & 45 & 1.4     & 0.5 & 78  \\
MWC 758(B) & A8 & 0   & 0.67 & 0.2  & 1.5 & 45 & 2       & 0.3 & 79  \\

\tablecomments{Col. (1) lists the spectral type used in the modeling; for all
other properties of the stars see Mannings \& Sargent (1997). Columns (2)--(7)
list the parameters determined from modeling. Overall parameters are (2) the
interstellar extinction to the system and (3) the fractional contribution of
the disk to the bolometric flux. The envelope parameters are (4) its overall
optical depth at visual and (5) the power of its density profile $y^{-p}$ ($y =
r/\Rsub$), which is terminated at $y = 1000$. \d The only exception is AB Aur,
whose envelope is modeled with a broken power law: $p = 2$ for $1 \le y \le
100$ and $p = 0$ for $100 < y \le 5000$. The disk parameters are its (6)
temperature (in K) and (7) 350 \mic\ optical depth at the outer edge. Derived
properties are (8) the disk observed diameter (in \arcsec) and (9) the envelope
inner radius.}

\enddata
\end{deluxetable}


\newpage

\figcaption{Fits to the SEDs of the MS sources with models comprised of
geometrically-thin optically-thick disks embedded in spherical dusty envelopes.
The data (de-reddened with the $A_V$ listed in Table 1) are marked with points,
LRS data (when available) with thick lines in the 8--24 \mic\ range. Each model
SED (full line) is the sum of the contributions of the envelope (dotted line)
and disk (dashed line) whose parameters are listed in Table 1. The
dashed--dotted line in each panel is the SED that a face-on disk would produce
if the envelope did not exist.}

\figcaption{Variation of temperature (top panel) and intensity at various
wavelengths with distance from the center along the apparent major axis of the
tilted disk for the AB Aur model (see Table 1; at the nominal distance to this
source, $0.1\arcsec \simeq 15$ AU).  In each panel the dotted line corresponds
to the envelope, the dashed line to the disk.  In the lower four panels,  the
overall intensity (full line) is the sum of the contributions of the two
components. Note the role reversal of the two intensity components between 10
and 100 \mic.}


\newpage

\begin{figure}
\centering \leavevmode \epsfxsize=\hsize \epsfbox[40 60  530 700]{fig1.ps}
\end{figure}

\newpage

\begin{figure}
\centering \leavevmode \epsfxsize=\hsize \epsfbox[50 85 540 730]{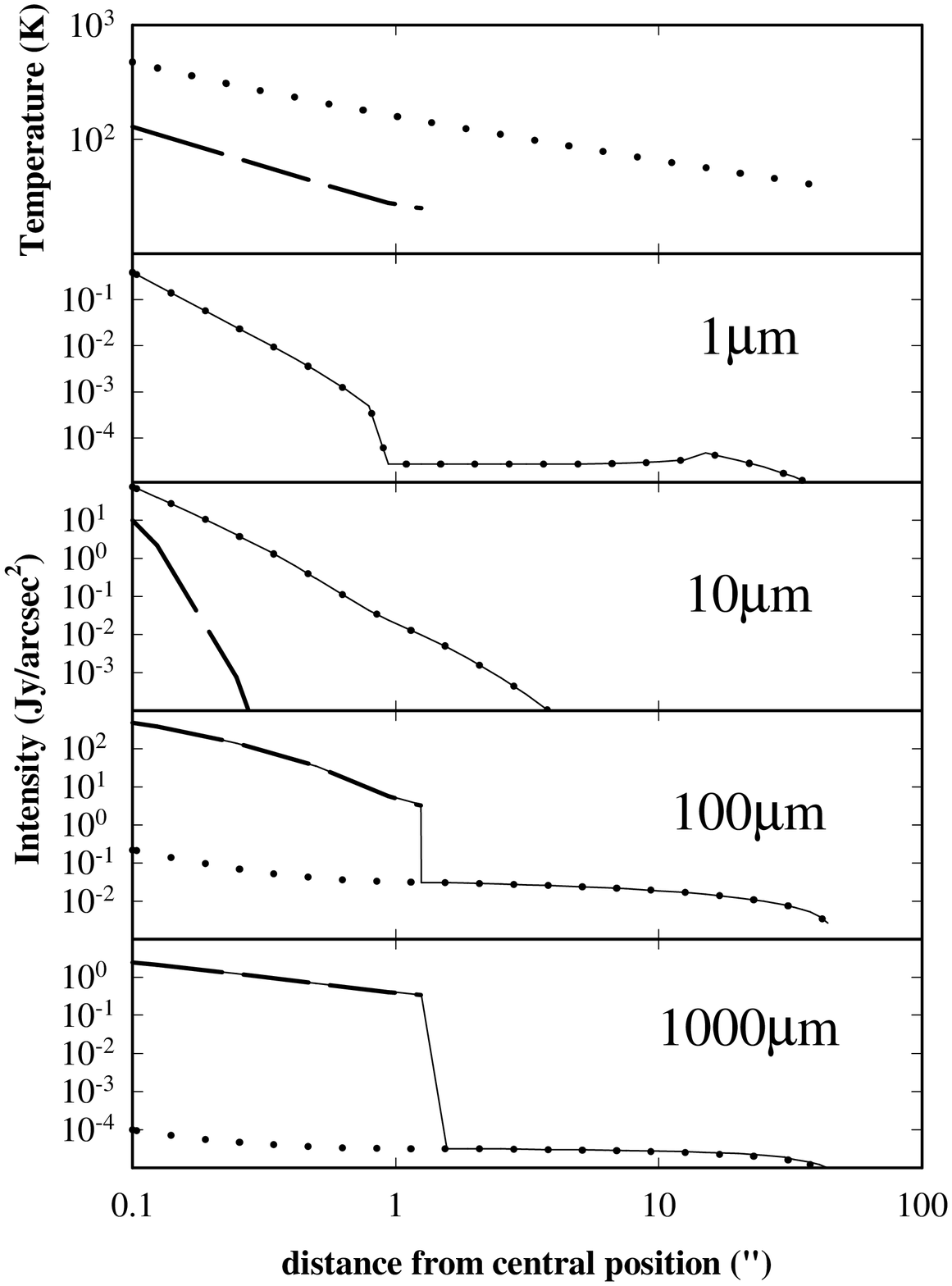}
\end{figure}

\end{document}